# Coherently controlled quantum features in a coupled interferometric scheme

Byoung S. Ham

Center for Photon Information Processing, School of Electrical Engineering and Computer Science, Gwangju Institute of Science and Technology

123 Chumdangwagi-ro, Buk-gu, Gwangju 61005, South Korea

bham@gist.ac.kr

**Abstract:**
Over the last several decades, entangled photon pairs generated by spontaneous parametric down conversion processes in both $\chi^{(2)}$ and $\chi^{(3)}$ nonlinear optical materials have been intensively studied for various quantum features such as Bell inequality violation and anticorrelation. In an interferometric scheme, anticorrelation results from photon bunching based on randomness when entangled photon pairs coincidently impinge on a beam splitter. Compared with post-measurement-based probabilistic confirmation, a coherence version has been recently proposed using the wave nature of photons. Here, the origin of quantum features in a coupled interferometric scheme is investigated using pure coherence optics. In addition, a deterministic method of entangled photon-pair generation is proposed for on-demand coherence control of quantum processing.

**Introduction**

Quantum entanglement [1] is the heart of quantum technologies such as quantum computing [2], quantum communications [3,4], and quantum sensing [5,6]. Although intensive research has been performed in both interferometric and noninterferometric schemes for quantum features such as a the Hong-Ou-Mandel (HOM) dip [7-9], photonic de Broglie wavelength (PBW) [10-13], Bell inequality violation [14-16], and Franson-type nonlocal correlation [17-20], the fundamental physics of entangled photon-pair generation itself has still been vailed in terms of probabilistic measurements via coincidence detection of coupled photon pairs. Thus, nondeterministic measurement-based quantum technologies have prevailed, resulting in an extreme inefficiency compared with their classical counterparts. The classical technologies are of course deterministic and macroscopic.

Recently, a novel method of deterministic quantum correlation has been proposed and demonstrated to unveil secretes of quantum entanglement for both HOM dip and PBW using the wave nature of photons [21-23]. As a result, the fundamental physics of quantum features has been found in the phase property of a coupled system, where the coupled system does not have to be confined by the Heisenberg's uncertainty principle. Based on this wave nature of photons, collective control of coherent photons becomes a great benefit for macroscopic quantum technologies compatible with the classical counterparts. Here, the fundamental physics of quantum correlation is investigated using the wave nature of photons to find the origin of quantum features demonstrated in an interferometric scheme [24]. For typical $\chi^{(2)}-$ generated entangled photon pairs, some misunderstanding regarding quantum correlation are pointed out not to criticize but to support the novelty of the wave nature of photons. Without violating quantum mechanics, a proper choice of photon property should depend on photon resources according to the wave-particle duality [25]. Finally, a coherence version of quantum feature generation is proposed for potential applications of deterministic and macroscopic quantum information processing.

Figure 1 shows a particular scheme of HOM-type quantum correlation in a coupled interferometric scheme, where entangled photon pairs are generated from spontaneous parametric down conversion (SPDC) processes in a $\chi^{(2)}$ nonlinear material [24]. Due to the spontaneous emission decay process, an initially given phase is randomly assigned to each photon pair, where each photon pair has also a random frequency detuning from the fixed half-frequency of the pump photon used for $\chi^{(2)}$. In related HOM-type experiments, a typical line shape observed by coincidence measurements shows a broad dip, whose decay is the inverse of the photon pairs' bandwidth. Unlike the theory in ref. [20] based on the wave nature of photons, however, $\lambda-$dependent $g^{(1)}$ correlation has never been observed. In the present paper, both reasons for the missing $g^{(1)}$ correlation in a single Mach-Zehnder interferometer (MZI) and the revival of $g^{(1)}$ correlation in a coupled MZI are investigated to unveil the physical origin of quantum features.

**Results**

For the analytic discussion as to why there is no $g^{(1)}$ correlation in a HOM dip, we start with a typical SPDC-generated entangled-photon system as shown in Fig. 1(a). Assuming there is a specific phase relation between the paired photons, signal $(E_S)$ and idler $(E_I)$, the basic equations for coincidence detection measurements can be derived using general matrix representations of pure coherence optics, where $\begin{bmatrix} E_\alpha \\ E_\beta \end{bmatrix} = [BS1][\zeta]\begin{bmatrix} E_I \\ E_S \end{bmatrix}$ and $\begin{bmatrix} E_A \\ E_B \end{bmatrix} = [BS2][\varphi][BS1][\zeta]\begin{bmatrix} E_I \\ E_S \end{bmatrix}$, $[BS2] = [BS1] = \frac{1}{\sqrt{2}}\begin{bmatrix} 1 & i \\ i & 1 \end{bmatrix}$, $[\zeta] = \begin{bmatrix} e^{i\zeta} & 0 \\ 0 & 1 \end{bmatrix}$, and $[\varphi] = \begin{bmatrix} 1 & 0 \\ 0 & e^{i\varphi} \end{bmatrix}$ [21]. Here, introduction of coherence optic is a choice matter without violation of quantum mechanics [25]. The j$^{th}$ input photon pair $E_{S_j}$ and $E_{I_j}$ are described with the wave nature property, where $E_{S_j} = E_0 e^{i(k_{S_j} r - 2\pi f_{S_j} t + \theta_{S_j})}$ and $E_{I_j} = E_0 e^{i(k_{I_j} r - 2\pi f_{I_j} t + \theta_{I_j})}$. The photon pair generation rate and bandwidth in SPDC can be controlled by adjusting a pump power and a spectral filter. In general, the detectable photon rate by a single photon detector module is far less than MHz. Considering a detection module speed larger than GHz, consecutive photon pairs are treated independently throughout the coincidence measurement process. The coherent property of each generated photon pair is determined by Heisenberg's uncertainty principle in terms of the energy-time relation: $\Delta f \Delta t \geq 1$. For a typical THz bandwidth $\Delta f$, the coherence time $\Delta t$ is in the order of ps. Compared with the corresponding coherence length $l_c \sim 100 \ \mu m$, the original wavelength $\lambda_0$ of the pump is far shorter than $l_c$. In other words, $g^{(1)}$ correlation is much sensitive than $g^{(2)}$ correlation.

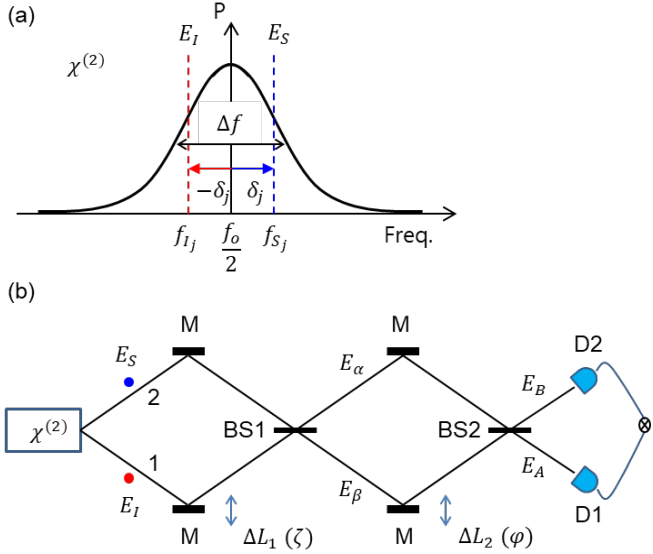

Fig. 1. Interferometric quantum feature generation. (a) A SPDC-based photon-pair bandwidth. (b) a SPDC-based coupled interferometric scheme. BS: beam splitter, D: detector, M: mirror. $\zeta = \frac{2\pi}{\lambda}\Delta L_1$; $\varphi = \frac{2\pi}{\lambda}\Delta L_2$; $\Delta$: bandwidth; $\delta_j$: random symmetric detuning of the j$^{th}$ entangled pair.

According to the energy conservation law, the signal and idler photons in each pair are symmetrically detuned by $\pm \delta_j$ from the half-frequency $(f_0/2)$ of the pump laser as shown in Fig. 1(a). Due to spontaneous emission processes, however, the frequencies $f_{S_j}$ and $f_{I_j}$ of the j$^{th}$ photon pair are random within the bandwidth $\Delta f$. Likewise, the initially given phases $\theta_{S_j}$ and $\theta_{I_j}$ are not determined, either. As analyzed, however, the difference phase $\delta \theta_j$ between $\theta_{S_j}$ and $\theta_{I_j}$ is fixed at $\pi/2$ [20]. This fact will be derived differently below based on Fig. 1(a). Figure 1(b) originats in ref. 24 and is used to understand important quantum features. The first (second) MZI is controlled by $\Delta L_1$ $(\Delta L_2)$, where $\zeta_j = \frac{2\pi}{\lambda_j}\Delta L_1$ $\left(\varphi_j = \frac{2\pi}{\lambda_j}\Delta L_2\right)$, and $\lambda_j$ is the j$^{th}$ photon's wavelength.

Regardless of nondegeneracy in $\chi^{(2)}$, all pairs are symmetrically detuned, whose corresponding phase difference is $\pm\delta_j\tau = \pm\zeta_j$, where $\tau$ is the relative delay between paired photons for measurements.

The coincidence measurements between output ports α and β on a beam splitter BS1 are for intensity correlation $g^{(2)}(\tau)$, where the j$^{th}$ output intensities are as follows (see Fig. S1 of the Supplementary Information):

$$I_\alpha^j(r,t) = I_0[1 + sin(\zeta_j')], \tag{1}$$
$$I_\beta^j(r,t) = I_0[1 - sin(\zeta_j')]. \tag{2}$$

Here, the phase $\zeta_j'$ is described as:

$$\zeta_j'(r,t) = \left(\frac{k_0}{2} - \delta k_j\right)\Delta L_1 - \left(\frac{\omega_0}{2} - \delta\omega_j\right)\tau + \delta\varphi_j - 2(\delta k_j r_s - \delta\omega_j t_s). \tag{3}$$

For all $\delta_j$-dependent photon pairs, $I_\alpha = \sum_j I_\alpha^j$ and $I_\beta = \sum_j I_\beta^j$. Equation (3) represents four different sources of the induced phase $\zeta_j'$. The first one $\left(\frac{k_0}{2}\Delta L_1\right)$ is a center frequency-related fundamental oscillation as a function of $\Delta L_1$: $2\lambda_0$-dependent fast oscillation. The second one $(\delta k_j \Delta L_1)$ is the detuning-caused slow oscillation, resulting in $\Delta f^{-1}(\tau)$-dependent decoherence. The third one $(\delta\varphi_j)$ is for a fixed relative phase $\pi/2$ between the signal and idler photons in each pair. The last one $(2\delta k_j r_s)$ is for $\Delta L_1$-independent frequency beating between the paired photons, resulting in a fixed phase. Due to the wide spectrum in Fig. 1(b), this beating results in a $\Delta f^{-1}(\tau)$-dependent wide envelope. Thus, equation (3) becomes a function of $\Delta L_1$ (or $\tau$) only. However, all $\delta f_j$-caused phase factors in equation (3) cancel each other out due to the $\pm\delta f_j$ distribution of all photon pairs except for the fixed $\delta\varphi_j$ at coincidence detection. Thus, the mean values of the output intensities are uniform, resulting in $\langle I_\alpha \rangle = \langle I_0 \rangle = I_0$ due to $\langle sin(\zeta_j')\rangle = 0$, where the signal and idler photons are interchangeable. This is the physical origin why there is no $g^{(1)}$ correlation in $g^{(2)}(\tau)$ in the first MZI. As analyzed for the second MZI below, this fact also becomes the physical origin as to how $g^{(1)}$ is retrieved in $g^{(2)}(\tau)$ as observed in ref. 24.

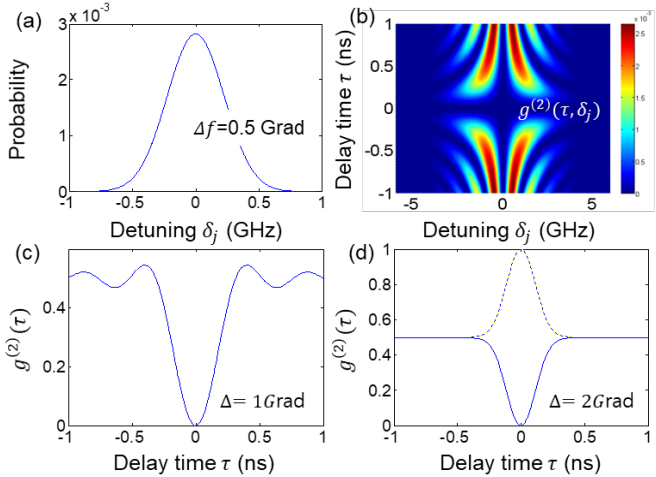

Fig. 2. Numerical simulations of intensity correlation in a typical HOM dip with $\delta\theta_j = \pm\pi/2$. (a) Photon distribution. (b) $\tau$ vs. $\delta_j$. (c) and (d) Sum $g^{(2)}(\tau)$ for all $\delta_j$ for different coverage $\Delta f$. Dotted: $\delta\theta_j = 0$.

According to the definition of intensity correlation $g_{\alpha\beta}^{(2)}(\tau) = \frac{\langle I_\alpha I_\beta \rangle}{\langle I_\alpha \rangle \langle I_\beta \rangle}$, the following equation results in:

$$g_{\alpha\beta_j}^{(2)}(\tau, \delta_j) = cos^2(\zeta_j'). \tag{4}$$

To satisfy anticorrelation of $g_{\alpha\beta_j}^{(2)}(\tau = 0, \delta_j) = 0$, in a SPDC-based HOM dip, $\delta\theta_j = \pm\pi/2$ must be satisfied for each photon pair [20]. If $\tau \neq 0$, equation (4) gradually decays and shows a typical HOM dip curve as a function of delay time $\tau_c$ ($= \Delta f^{-1}$), where decay time $\tau_c$ in $g^{(2)}(\tau)[=\sum_j g_j^{(2)}(\tau, \delta_j)]$ is preset according to the inverse of the SPDC-generated photon bandwidth $\Delta f$ as shown in Fig. 2.

Figure 2 shows numerical calculations for equation (4). Figure 2(a) shows the Gaussian distribution of SPDC-generated photon pairs with the bandwidth of $\Delta f = 0.5 \times 10^9$ radians. According to Fig. 1, the j$^{th}$ photon pair has different detuning at $\delta f_j$, whose corresponding $g^{(2)}(\tau, \delta f_j)$ is shown in Fig. 2(b). In equation (4), the j$^{th}$ photon pair must have different $g^{(2)}(\tau)$ only due to the detuning dependent $\zeta_j'$. By definition, $g^{(2)}(\tau)$ is obtained via averaging all $\delta f_j$−dependent coincidence measurements for a fixed τ. As shown in Figs. 2(c) and (d), maximum $g^{(2)}(\tau)$ is bound to $g^{(2)}(\tau) = 0.5$, where $g^{(2)}(\tau) = 0.5$ is a classical lower bound [20]. This upper limit of $g^{(2)}(\tau) = 0.5$ strongly supports the nonclassical phenomenon of entangled photon pairs [24]. If all of the spectral photon pairs are not fully covered for measurements, there is a wiggle in $g^{(2)}(\tau)$ as shown in Fig. 2(c). This wiggle is due to incomplete coherence washout in the summation process [24]. Disappearance of the $g^{(1)}$ fringe in a HOM dip is not due to the measurement process or artifacts, but instead due to the inherent properties of the symmetrically detuned photon pairs in SPDC. If there is no relative phase between signal and idler photons, then there is no nonclassical feature in $g^{(2)}(\tau)$ as indicated by the dotted curve in Fig. 2(d). If the relative phase $\delta\theta_j$ is random for all pairs, $g^{(2)}(\tau) = 1/2$ regardless of τ, representing the property of individual particle ensemble [20].

In the second MZI in Fig. 1(b), the $\Delta L_2$ effect can be classified for bunched photons only on BS1 if $\Delta L_1 \sim 0$. According to equation (3), all other terms become zero except for $\delta\varphi_j$, which is $\pi/2$ for all j. Here, it should be noted that the bunched photons in each path of the MZI are composed of signal and idler photon pairs, whose detuning is exactly opposite across the center frequency $f_0/2$. Thus, whenever a nonzero $\Delta L_2$ occurs, the detuning $\delta f_j$−caused phase terms in equation (3) are cancelled out automatically due to the $+/-$ relation in $\delta f_j$. As a result, only the original $2\lambda_0$−dependent fast oscillation survives in the output fields. This is the unspoken secretes in the SPDC-based $g^{(1)}$ features observed in ref. 24 for $g^{(2)}$ measurements.

In the second MZI of Fig. 1(b), the following amplitude relations are obtained for the final outputs $E_A$ and $E_B$:

$$\begin{bmatrix} E_A \\ E_B \end{bmatrix}_j = \frac{1}{2} \begin{bmatrix} 1 - e^{i\varphi_j} & ie^{i\zeta_j'}(1 + e^{i\varphi_j}) \\ i(1 + e^{i\varphi_j}) & -e^{i\zeta_j'}(1 - e^{i\varphi_j}) \end{bmatrix} \begin{bmatrix} E_S \\ E_I \end{bmatrix}_j. \quad (5)$$

From equation (5), the corresponding intensities are as follows (see Fig. S2 of the Supplementary Information):

$$I_A^j = I_0(1 - \cos\zeta_j' \sin\varphi_j), \quad (6)$$
$$I_B^j = I_0(1 + \cos\zeta_j' \sin\varphi_j). \quad (7)$$

The anticorrelation condition $\zeta_j' = \pm\pi/2$ in equation (3), however, results in independence of $\varphi_j$. If $\zeta_j' = 0$, $I_A^j = I_0(1 - \sin\varphi_j)$ and $I_B^j = I_0(1 + \sin\varphi_j)$ are obtained. In this case, however, the photon bunching or anticorrelation in equations (3) and (4) is violated, resulting in the classical feature of $g^{(2)}_{\alpha\beta_j}(\tau, \delta_j) = 1$ from equation (4) (see Fig. S3 of the Supplementary Information). Although the normalized coincidence detection measurement becomes $R_{AB}{}^j = \frac{1}{2}(1 + \cos 2\varphi_j)$, $g^{(2)}_{AB_j}(\tau_\zeta, \tau_\varphi, \delta_j) = 1$ shows a classical feature. In fact, the $\cos 2\varphi_j$ modulation term in $R_{AB}$ is a typical classical feature of the intensity product from a single MZI. In other words, satisfying $g^{(2)}_{\alpha\beta_j}(\tau, \delta_j) = 0$ for photon bunching violates $R_{AB}{}^j = \frac{1}{2}(1 + \cos 2\varphi_j)$ (see Fig. S3 of the Supplementary Information). Thus, the observations of $\cos 2\varphi_j$ modulation in ref. 24 are not quantum but classical as shown in Fig. 3(d). To be quantum, both MZI paths must have bunched photons within coherence time as demonstrated [12,13]. To work with $N \geq 4$ for PBW, an inter-MZI superposition scheme can be a quantum solution as proposed [21,22] and demonstrated [23].

Figure 3 shows a coherence version of the entangled photon-pair generation comparable with Fig. 1. Because MZI works for either a single photon or coherence light equivalently [26], the results have no difference for the photon characteristics. The photons propagating along different paths of MZI 1 is strongly coupled by the relative phase of $\pi/2$ created from the first BS, regardless of the input photon's wavelength [27]. The matrix representations for Fig. 3(a) are as follows without considering $\Delta L_1$: $I_\alpha = \frac{I_0}{2}(1 - \cos\zeta)$, $I_\beta = \frac{I_0}{2}(1 + \cos\zeta)$,

$I_A = \frac{I_0}{2}[1 - sin\varphi sin\zeta]$, and $I_B = \frac{I_0}{2}[1 + sin\varphi sin\zeta]$ (see Fig. S4 of the Supplementary Information). Using an acousto-optic modulator (AOM) driven by an rf pulse generator with an rf frequency of $f_{rf}$, the role of $\delta f_j$ −caused random phases in Fig. 1 can be satisfied by a 50% duty cycle of AOM between 0 and $f_{rf}$, as shown in Fig. 3(b). In other words, the zeroth (original $f_0$) and first-order ($f_0 + f_{rf}T$) diffracted light pulses are used, where T is the rf pulse duration. If $2\pi f_{rf}T = \pi$, the output direction is reversed. Thus, the average of each output intensity becomes uniform, $\langle I_\alpha \rangle = \langle I_\beta \rangle = \langle I_A \rangle = \langle I_B \rangle = I_0$, satisfying randomness. Including the $\Delta L_1$ effect in $\zeta$, the revised output intensities are as follows:

$I_\alpha = \frac{I_0}{2}(1 - cos\zeta')$, (8)

$I_\beta = \frac{I_0}{2}(1 + cos\zeta')$, (9)

$I_A = \frac{I_0}{2}[1 - sin\varphi sin\zeta']$, (10)

$I_B = \frac{I_0}{2}[1 + sin\varphi sin\zeta']$, (11)

where $\zeta' = \zeta + k\Delta L_1 (2\pi f_{rf}T)$.

Figure 3(c) shows numerical calculations for equations (8)-(11) (see also Figs. S4 and S5 of the Supplementary Information). For $\zeta' = \zeta + \pi/2$, equations (8)-(11) are rewritten as $I_\alpha = \frac{I_0}{2}(1 + sin\zeta)$, $I_\beta = \frac{I_0}{2}(1 - sin\zeta)$, $I_A = \frac{I_0}{2}[1 + sin\varphi cos(\zeta)]$, and $I_B = \frac{I_0}{2}[1 - sin\varphi cos(\zeta)]$. The normalized intensity product $R_{ij}$ between $I_i$ and $I_j$ is the same as $g_{\alpha\beta}^{(2)}(\zeta) = \frac{1}{2}(1 - cos2\zeta)$ for MZI 1 and $g_{AB}^{(2)}(\varphi) = \frac{1}{2}(1 - sin^2\varphi sin^2\zeta)$ for MZI 2 due to the randomness by AOM. To satisfy the anticorrelation condition for $g_{\alpha\beta}^{(2)}(\zeta)$, $\zeta = \pm\pi/2$ is obtained as shown in the top panels of Fig. 3(c). For the same conditions of $\zeta = \pm\pi/2$, however, there is no way to satisfy the quantum feature between $I_A$ and $I_B$, unless $\Delta L_1$ is changed. For $R_{AB} = 0$, $\zeta = \pm n\pi$ must be satisfied as shown in the bottom panels, where n=0,1,2… As analyzed in Fig. 2, this also proves the violation of quantum feature analysis in ref. 24. In a short summary, the correct condition for the quantum feature generation in Fig. 3(a) for the final outputs is to break the anticorrelation condition in ζ. Neither way, the PBW cannot be possible in the directly coupled MZI scheme due to this reason, where Fig. 3(c) is just for the diffraction limit of the Rayleigh criterion in the intensity product: $R_{AB} = (1 + cos2\varphi)/2$. As presented elsewhere, such PBW can be achieved by CBW via path superposition [28]. For this, an intermediate dummy MZI must be inserted between two MZIs in Fig. 3(a).

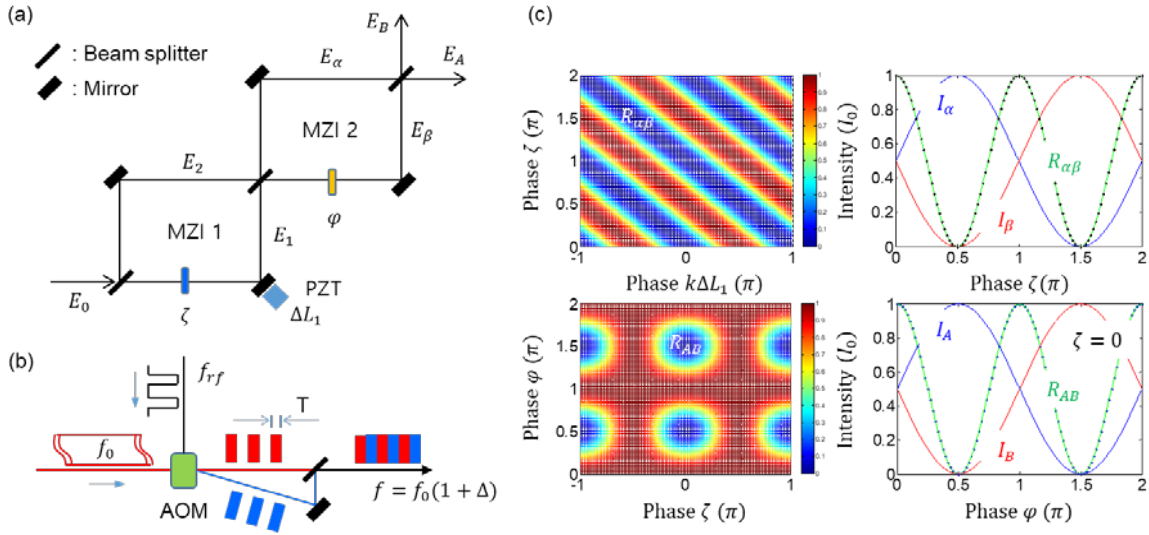

Fig. 3. Schematic of deterministic entangled photon-pair generations. (a) A coupled MZI structure. (b) Basis randomness for ζ (0; π). (c) Numerical calculations for $R_{ij} = I_iI_j * 4$ at $k\Delta L_1 = \frac{\pi}{2}$. (Top row) For $I_\alpha$ and $I_\beta$. (bottom row) For $I_A$ and $I_B$. $I_i$ and $I_j$ are interchangeable on behalf of AOM.

**Conclusion**

In conclusion, the quantum features of anticorrelation and PBW were analyzed in a directly coupled MZI system using pure coherence optics, where SPDC-generated symmetrically distributed entangled photon pairs played an essential role in both $g^{(1)}$ disappearance in the first MZI and $g^{(1)}$ retrieval in the second MZI. Based on the $\chi^{(2)}$ − generaged entangled photon-pair distribution, the relative $\pi/2$ phase difference between all paired photons was derived as an essential condition for anticorrelation. Moreover, anticorrelation condition in the first MZI violated quantum feature generation conditions in the second MZI. In other words, satisfying the anticorrelation in one MZI resulted in destruction of quantum features in the other MZI. By this reason, PBW could not be generated from the directly coupled MZI system. Instead, quantum superposition between MZIs is the solution of creation of PBW as presented in refs. 21-23 and 28. Finally, a deterministic coherence version of entangled light pair generation was proposed and analyzed using pure coherence optics applicable to both single photons and coherent light without violation of quantum mechanics.

Acknowledgment

BSH acknowledges that this work was supported by GIST via GRI 2021.